\documentclass[aps,prb,twocolumn,reprint,superscriptaddress,showkeys,floatfix,amsmath,amssymb]{revtex4-1}

\usepackage{natbib}
\usepackage[utf8]{inputenc}
\usepackage{epsfig}
\usepackage{color}
\usepackage{booktabs}
\usepackage{url}
\usepackage{graphicx}
\usepackage{csquotes}
\usepackage{xcolor, soul}
\usepackage{tabularx}
\usepackage{dcolumn}
\usepackage{bm}
\usepackage{amsmath}
\newcommand{\rtcom}[1]{\hl{}}{}
\usepackage{natbib}

\usepackage{hyperref}

\begin{document}
\preprint{APS/123-QED}

\title{Cu\texorpdfstring{$_2$}{2}XSiS\texorpdfstring{$_4$}{4} (X = Ge, Sn, and Pb) materials for solar-cell applications: A DFT+SCAPS-1D simulation}%

\author{H. Laltlanmawii}
\affiliation{Department of Physics, Mizoram University, Aizawl-796004, India}
\affiliation{Physical Sciences Research Center (PSRC), Department of Physics, Pachhunga University College,  Aizawl-796001, India}
\author{L. Celestine}
\affiliation{Department of Physics, Mizoram University, Aizawl-796004, India}
\affiliation{Physical Sciences Research Center (PSRC), Department of Physics, Pachhunga University College,  Aizawl-796001, India}
\author{R. Zosiamliana}
\affiliation{Department of Physics, Mizoram University, Aizawl-796004, India}
\affiliation{Physical Sciences Research Center (PSRC), Department of Physics, Pachhunga University College,  Aizawl-796001, India}
\author{B. Chettri}
\affiliation{Department of Physics, Mizoram University, Aizawl-796004, India}
\author{S. Bhattarai}
\affiliation{Technology Innovation and Hub, Indian Institute of Technology Guwahati, Guwahati, Assam, 792103, India}
\author{K. C. Bhamu}
\affiliation{Department of Physics, SLAS, Mody University of Science and Technology, Lakshmangarh, Rajasthan, 332311 India}
\author{D. P. Rai}
\email[D. P. Rai: ]{ dibyaprakashrai@gmail.com}
\affiliation{Department of Physics, Mizoram University, Aizawl-796004, India}

\date{\today}

\begin{abstract} By means of the first-principles density functional theory (DFT), I$_2$-II-IV-VI$_4$ type Cu-based quaternary chalcogenides Cu$_2$XSiS$_4$ (X = Ge, Sn, and Pb) have been thoroughly investigated. We report the study of Ge and Sn substitution in the divalent cation site for their potential applications in photovoltaics for the first time. The structural, electronic, optical, and mechanical properties have been calculated. The structural and thermal stability is verified by calculating the elastic constants, formation energy and total potential energy at 300 K from the ab-initio molecular dynamics (MD) simulation.
The compounds under our investigation exhibited an indirect band gap in the range of 1.0--1.56 eV, suitable for energy harvesting by trapping the sunlight. The presence of absorption peaks within the visible region complements their potential in photovoltaic applications.  For further validation, we have designed a model of a heterostructure (FTO/TiO$_2$/Cu$_2$XSiS$_4$/CuO/Au) solar cell, and a numerical simulation has been performed by solving the Poisson equation and continuity equations to obtain the I-V characteristic by using SCAPS-1D. All the inputs needed for solar-cell simulation in SCAPS-1D have been taken from the DFT results. The corresponding Power Conversion Efficiency (PCE) is denoted by $\eta$\% and their respective values for X=Ge, Sn and Pb are 23.46\%, 23.29\% and 22.60\%, at room temperature. The Ge-based system exhibits the highest $\eta$\%, owing to its band gap value in the visible range of the solar spectrum. Thus, we report that Ge-based compounds may act as a promising absorber layer in heterostructure solar-cell applications.

\end{abstract}

\maketitle


\section{Introduction} 
As the global demand for green and surplus energy sources is growing over time, solar energy harvesting has emerged as a possible alternative alongside other renewable resources such as wind \cite{Su2025}, hydroelectricity \cite{Ge2024}, geothermal \cite{Anya2025}, and biomass energy \cite{Ahmadipour2025, Ratnesh2025}. Solar energy is clean and pollution-free, as opposed to traditional energy sources such as fossil fuels, which affect the environment by emitting greenhouse gases and causing climate change, as well as air and water pollution \cite{Al-Ezzi2022}. However, due to the inherent variability of sunlight, an energy storage device is required to perpetually use solar electricity \cite{Sajitha2024}. This can be performed by photovoltaics (PV), a technology that converts sunlight directly into electrical energy using semiconductor materials \cite{Dambhare2021}. First-generation silicon solar cells continue to dominate commercial PV technology, with a power conversion efficiency of 26\% \cite{Liang2025}. Thin Film Solar Cells (TFSCs) have also emerged as a promising candidate; nevertheless, the market shares of CdTe are 5\% and Copper-Indium-Gallium-Selenide (CIGS) solar cells are just 2\%, which is significantly lower than those of Si-based solar cells of over 90\% \cite{Maalouf2023}. This can be attributed to their lower power conversion efficiency (PCE) (22\% for CdTe and 23.4\% for CIGS \cite{Nakamura2019}), Cd toxicity \cite{Tai2025}, and the scarcity of elements such as Indium, Gallium, and Tellurium, which limit their potential for low-cost and high-volume manufacturing \cite{Paul2024}.

\par Generally, the development of a new earth-abundant and non-toxic compound to further improve the performance and cost characteristics of TFSCs consists of expanding the semiconductor through the cross-substitution technique. This involves the replacement of an element with two other elements while maintaining the octet rule and a consistent atom ratio. \cite{Hossain2025, Ghorbani2020} Due to the scarcity of In and Ga, a cross-substitution aimed at the trivalent metal cation site by Zn(II) and Sn(IV) in Cu(In/Ga)S$_2$ (CIGS) results in quaternary chalcogenides Cu$_2$ZnSnS$_4$ known as Kesterite or CZTS. \cite{Saparov2022} CZTS has gained significant interest in the PV world as they are made up of environmentally friendly elements, along with a high absorption coefficient of 10$^{-4}$ cm$^{-1}$ and a direct band gap (1.0 - 1.6 eV), which is favourable for absorbing solar radiation.\cite{Fatima2025} While the Shockley-Queisser limit for Kesterite is 31\%, the power conversion efficiency remains at 14.9\%, which is significantly lower compared to the parent CIGS.\cite{Yao2025} The main technical barrier is the complexity of the crystal chemistry: the fact that CZTS combine four different elements (Cu, Zn, Sn, and S/Se) leads to cation disorder, \cite{Bosson2017} phase instability,\cite{Gupta2019} and voltage losses.\cite{Yang2017} These are difficult to design without sophisticated synthesis techniques and a deeper theoretical grasp of defect physics.\cite{Chennangod2025}

\par CZTS is thermally stable with a larger band gap than its Cu$_2$ZnSnSe$_4$ (CZTSe) and  Cu$_2$ZnSn(S,Se)$_4$ (CZTSSe) counterparts. On the other hand, CZTSe has a lower band gap and which is better suited for infrared absorption with a lower open-circuit voltage. CZTSSe is a combination of sulfur and selenide in the anionic composition, and they offer band gap tuning by varying the S/Se ratio, balancing voltage and current.\cite{Ji2013, Grossberg2019, Simya2015} Challenges like cation disorder and defect passivation remain crucial for all three types of Kesterite.\cite{Wei2023, Dale2009, Zhou2013} For improving the device performance, a theoretical solar cell module has been designed and simulated using the SCAPS-1D program by many researchers.\cite{KHAN2026, Singh2025, Ouarie2023}  To achieve a realistic PV response, a variation in minority carrier lifetime, defect concentration, interface, and layer thickness can be introduced to optimize device performance.\cite{Kukreti2021, GUERROUM2023}  The conventionally used CdS buffer layer has non-optimal band alignment with the absorber layer in kesterite-based solar cells, in addition to being toxic, which limits their efficiency. To address this issue, Tseberlidis \textit{et al}. introduced TiO$_2$ as a suitable buffer layer for solar cell heterostructure.\cite{Tseberlidis2023} Recently, Lofty \textit{et al}. has reported the efficiency above 30\% by using TiO$_2$ as electron transport layer (ETL), CuO as hole transport layer, and optimized different parameters such as thickness, carrier concentration, band gap etc.\cite{Lofty2025} Employing MoO$_x$ as back surface and Au as back contact have also been found to enhanced PCE.\cite{ABDELKADIR2023, Tarekuzzaman2025} Arockiadoss \textit{et al}. shows that oxide (Cu$_2$O) enhances performance with efficiency reaching about 18.48\% when used as a hole transport layer.\cite{AROCKIADOSS2025} Aside from the absorber layer, kesterite materials are also found to be an excellent HTL material in solar cells.\cite{DIXIT2023, Mahsar2025} Jakalase \textit{et al}. utilised CZTSe, CZTSSe, CNTS and CFTS as HTL in Quantum dot-sensitized solar cells (QDSSCs). With a PCE of 25.86\%, the CFTS material was found to outperform the others.\cite{Jakalase2016} Et‑taya \textit{et al}. introduced a double absorber CZTSSe solar cell by incorporating CIGS as the bottom absorber layer. This device model yields a remarkably high efficiency of 45.40\%.\cite{Et-taya2025} 

\par Researchers have investigated a variety of cation substitution methods to overcome the inherent constraints of CZTS. For example, Cu replaced by Ag minimises the Cu-Zn disorder while also lowering the deep defect levels.\cite{Das2022, Gong2021} Zn-substituted with different divalent metals can alter the chemical bonding and stabilise the structure, depending on the substituent element. The substitution of Zn with Cd is most widespread due to its strong lattice match and ability to inhibit Cu$_{Cd}$-related point defects and defect clusters, resulting in a few band-tailing states.\cite{Zhang2024} However, Cd is toxic, thus other materials such as Mn and Mg have been explored for sustainable Kesterites as they are eco-friendly.\cite{Malik2021, Abdullah2024, Pansuriya2022}Cu$_2$CoSnS$_4$ exhibits a wide band gap whose value can go up to 1.7 eV and high optical absorption $>$10$^{-4}$ cm$^{-1}$ in the visible spectrum, which covers the majority of the solar radiation spectrum, making it a good candidate for absorber layer material.\cite{Harrathi2023, Oubakalla2022, Guesmi2023} Other divalent cation substituent elements, such as Fe, Ni, Ba, etc., are also being explored.
\cite{Drissi2025, Ghemud2023, Khatun2023, Jyoti2024, Kadari2024, Elsaeedy2019} When the tetravalent cation site is occupied by Ge, the band gap is found to be significantly widened compared to the parent CZTS, and it exhibits moderate hole mobility.\cite{Cheng2023} Si substitution on the Sn-site is also found to enhance the PCE when used as an absorber layer in a solar cell.\cite{Immanuel2023} 

\par In this paper, following the cross-substitution trend as mentioned above, we extend the designed concept of I$_2$-II-IV-VI$_4$ type quaternary chalcogenides with the general formula Cu$_2$XSiS$_4$ where X=Ge, Sn, and Pb. The compound Cu$_2$PbSiS$_4$ has been successfully synthesized by Olekseyuk \textit{et al} using the powder synthesis technique.\cite{Olekseyuk2005} It crystallized in trigonal structure (space group P3$_2$21) and exhibits promising properties for PV application. However, due to the toxicity of Pb, we considered an environmentally friendly alternative, Ge and Sn, as potential replacements at the X-site. Although Ge and Sn are typically tetravalent (stable in the +4 oxidation state), they can also exist in the +2 oxidation state.\cite{Gill2022} Therefore, we explore them as divalent substitutional elements at the X-site. To date, there is no reported literature on the substitution of Ge and Sn at divalent metal cation sites in these compounds. Consequently, this study aims to fill this gap by evaluating the structural, electronic, and optical properties of these proposed materials. Therefore, we perform first-principles calculations based on density functional theory (DFT) to determine their stability and potential as PV materials. To predict the device performance and validate their suitability as an absorber layer, we have performed a simulation of FTO/TiO$_2$/Cu$_2$XSiS$_4$/CuO/Au heterojunction solar cell through SCAPS-1D software. The PV parameters, such as open-circuit voltage, short-circuit current, fill factor, and PCE, were studied under standard illumination (AM 1.5 G 1000 W/m2).

\section{Computational Details} 
All calculations in this work were carried out using QuantumATK Q-2019.12, which employs the linear combination of atomic orbitals (LCAO) approach within the framework of density functional theory (DFT) \cite{Nityananda2017a, Smidstrup2017b, Schlipf2015}. For all three compounds, the electron-ion interactions were modeled using the generalized gradient 
approximation (GGA) with the Perdew-Burke-Ernzerhof (PBE) exchange-correlation functional. \cite{Perdew1996c} Structural optimization involved relaxing both atomic locations and unit cell parameters. To ensure accurate convergence of total energy and optimum geometry, a strict convergence criteria for the Hellmann-Feynman force convergence threshold of 0.02 eV \r{A}$^{-1}$ and stress tolerance of 0.0006 eV \r{A}$^{-3}$ were used. For Brillouin zone integration, a Monkhorst-Pack \textit{k}-point mesh of 8$\times$8$\times$8 was employed for structural optimization. \cite{Monkhorst1976a} A medium basis set was applied for all atoms (Cu, Si, S, Ge, Sn, and Pb) using Pseudo-Dojo pseudopotentials, with a density mesh cut-off at 105 Hartree. \cite{VanSetten2018a} For the calculation of the electronic and optical properties, a denser k-point mesh of 12$\times$12$\times$12 was employed, using the same convergence parameters as used for the total energy and the forces. For designing and simulation of heterojunction solar cell, we have used SCAPS-1D.\cite{Burgelman2000, Niemegeers1998, Degrave2003, Verschraegen2007, Lam2020, Bouri2024, Thomas2021, Heera2025} For detail architecture of the software, see Section \ref{scaps}. 

\section{Results and Discussions}
\subsection{Structural properties} 
\begin{figure}[hbt!]
	\centering
	\includegraphics[height= 6cm]{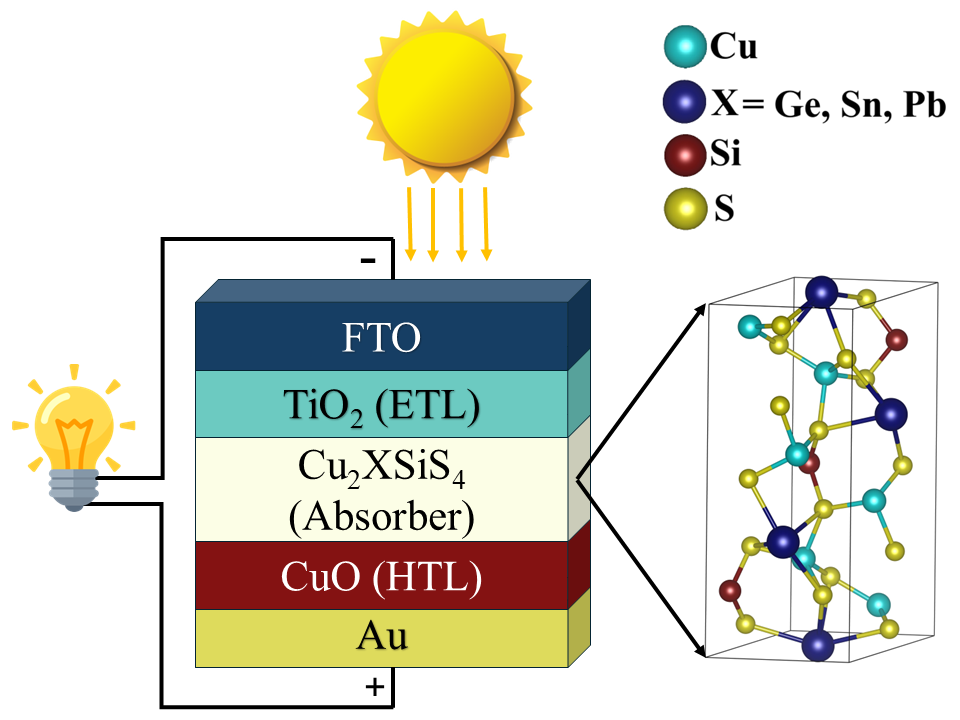}
	\caption{Schematic diagram of FTO/TiO$_2$/Cu$_2$XSiS$_4$/CuO/Au heterojunction solar cell}
	\label{Fig.1 Structures}
\end{figure}

In this study, we investigated the optimized structures of quaternary chalcogenides Cu$_2$XSiS$_4$, where X= Ge, Sn, and Pb. All compounds adopt a trigonal structure under the hexagonal system, specifically crystallizes in the P3$_2$21 (154) space group symmetry (see Figure.\ref{Fig.1 Structures}). The structural modelling and set-up of these compounds were performed using the 3D Visualisation for Electronic and Structural Analysis (VESTA) program.\cite{Momma2008} The structural data for Cu$_2$PbSiS$_4$ was obtained from the open-access online materials database "Materials project".\cite{Jain2013} However, for Cu$_2$GeSiS$_4$ and Cu$_2$SnSiS$_4$, no prior computational data were available. The optimized lattice constants \textit{a=b} and \textit{c}, unit cell volumes (V), band gap (E$_g$) and the predicted formation energies (E$^f$) along with available experimental data are summarized in Table \ref{Table 1}. The in-plane lattice parameter \textit{a} increases from 6.10\r{A} (Pb) to 6.12\r{A} (Ge), showing lattice expansion as the ionic radii dropped; however, the out-of-plane parameter 'c' exhibited contradictory behaviour: with Sn being the largest (15.29 \r{A}), followed by Pb (15.17\r{A}), and Se (15.13\r{A}). This indicates anisotropic lattice distortion, which occurs when each substituent introduces different ionic sizes and bonding conditions, resulting in distinct strain responses along the c-axis.\cite{Xiao2025} Interestingly, despite a drop in lattice parameters, the overall unit cell volume grew gradually from Ge (490.7 \r{A}$^3$) to Sn (494.8 \r{A}$^3$) and Pb (495.7 \r{A}$^3$). The decreasing ionic radii of replacing cations cause varied degrees of compressive strain within the lattice, supporting this surprising tendency. The observed anisotropy and volume contraction may impact the material's behaviour. The theoretical lattice parameters and unit cell volume align well with experimental results for Cu$_2$PbSiS$_4$.\cite{Olekseyuk2005} The slight discrepancies observed for the theoretical calculations may arise from the choice of exchange-correlation functional and inherent limitations in DFT approximations.\cite{Cohen2012}

\par We evaluated the stability of the studied compounds by calculating their formation energy using the equation below \cite{Celestine2025} 
\begin{equation}
	E^f = \frac{E_{total}-(6E_{Cu}+3E_{X}+3E_{Si}+12E_{S})}{N}
	\label{Eq 1}
\end{equation}
where E$_{total}$ is the total energy, E$_{Cu}$ is the total energy of Copper atom, E$_X$ are the energies of divalent atoms (Ge, Sn, and Pb), E$_{Si}$ is the total energy of Si, and E$_{S}$ is the total energy of S, and N is the number of atoms in the compound. The formation energy of a compound is the amount of energy required to combine its constituent elements in their standard states, reflecting their thermodynamic stability and likelihood of formation. A negative formation energy indicates that the compound is more stable than the individual constituents.\cite{Baryshev2015} This means that energy is released when the chemical forms, making it a positive process. In contrast, a positive formation energy indicates that the compound is less stable and requires energy to form.\cite{Lu2025} From Table \ref{Table 1}, it can be seen that all compounds are stable, as their formation energies are negative. Cu$_{2}$PbSiS$_4$ has the lowest E$^f$ value, indicating the highest internal energy stability among the three compounds.
\begin{table}[hbt!]
	\small
	\caption{Calculated lattice constants \textit{a} and \textit{c} in \r{A}, volumes (V) in \r{A}$^3$, band gap (E$_g$ in eV), formation energy (E$^f$ in eV) and effective masses of electron/hole (m$_e$/m$_h$ in m$_o$ rest mass of electron) for Cu$_2$XSiS$_4$ (X = Ge, Sn, Pb).}
	\label{Table 1}\renewcommand{\arraystretch}{1.0}

	\begin{tabular}{cccccccc}
		\hline
		X & (a,c)& V&&  E$_g$&   E$_f$ & m$_e$/m$_h$  \\
		\hline
Ge & 6.12, 15.13 & 490.7 && 1.08 & -1.01&\\
Sn & 6.11, 15.29 & 494.8 && 1.17 & -1.07&\\
Pb & 6.10, 15.17 & 495.7 && 1.56 & -1.10&\\
Pb$^a$ & 6.05, 15.17 & 482.8 && (1.69$^*$,0.91) & &\\
		\hline
	\end{tabular} \\
	Note: $a$\cite{Nhalil2018} and "*" indicates experimental value 
\end{table}
\par To validate the thermal stability of the proposed compounds, molecular dynamics (MD) simulations were performed within the NVT canonical ensemble under constant temperature by using QuantumATK's Noose-Hover (NH) thermostat.\cite{Martyna1992} The standard NH thermostat is formulated based on the system temperature controlled by an external system.\cite{Chen2011} An NVT simulation can accurately model the structure stability of materials whose volume variation with temperature is small. The temperature control set-up using NH thermostat is ideal for investigating a material's phase response to changes in temperature.\cite{Goncalves1992} 
\par
\begin{figure}[hbt!]
	\centering
	\includegraphics[height= 7.5cm]{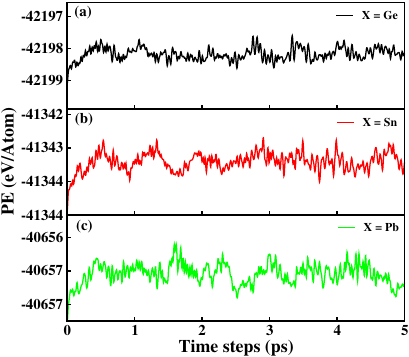}
	\caption{Molecular Dynamics (MD) simulations of Cu$_2$XSiS$_4$ (X = Ge, Sn, Pb)}
	\label{Fig.2 MD}
\end{figure}

 The evolution of potential energies in 5 ps time steps was monitored to assess the thermodynamic stability for the studied compounds,                      as illustrated in Figure \ref{Fig.2 MD}. We have employed the NVT-based canonical ensemble, and during these simulations, the temperature is kept constant at 300K. The number of particles, volumes, and temperature were kept constant for comprehensible results in the evolution of energies. From Figure \ref{Fig.2 MD}, an initial transient phase followed by fluctuations around a stable average can be observed, which indicates that each structure attained a thermal equilibrium; thus, we can assume that the studied compounds are thermally stable.
\subsection{Electronic properties} 
\begin{figure*}
    \centering
    \includegraphics[height=4.8cm]{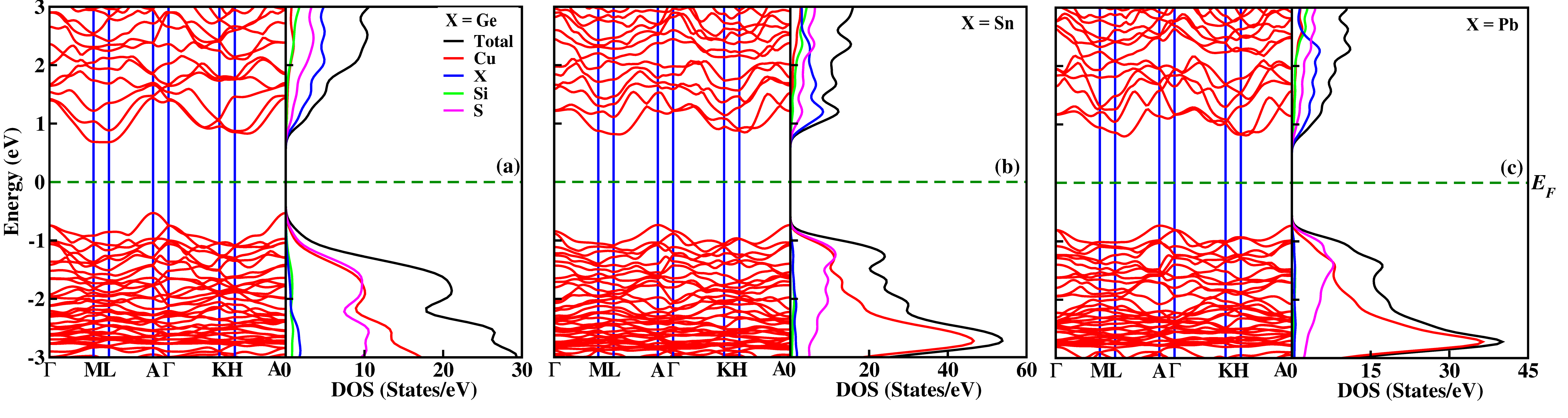}
    \caption{Band structures and Density of States for (a) X = Ge (b) X = Sn (c) X = Pb}
    \label{Fig.3}
\end{figure*}

Electronic properties are fundamental in understanding the behaviour of a material for a wide range of applications, including optoelectronic and photovoltaic devices.\cite{Lavrentyev2019} The electronic band structure and Density of States (DOS) were estimated by employing the GGA exchange functional to provide the electronic properties. All the examined compounds exhibit indirect band gaps and demonstrate semi-conductive nature as shown in Figure \ref{Fig.3}. Band structure is crucial for understanding physical features, including absorbance spectra, charge density distribution, and electron-hole mobility in photovoltaic materials.\cite{Chen2009} We evaluated the band structures with the energy band range of -3 to +3 eV and the zero-point energy as the Fermi energy (E$_f$). The valence band maximum (VBM) is located on the high symmetry A-points, and the conduction band minimum (CBM) is situated on the high symmetry L-points in the first Brillouin zone for all the structures, confirming the indirect band gap with semi-conductive nature for all our examined compounds. For  X = Ge, Sn and Pb, our obtained E$_g$ values are 1.08 eV, 1.17 eV, and 1.56 eV, which are all considered to be in a suitable range for solar cells and other photovoltaic applications. Furthermore, we have observed the band gap of X = Pb aligned well with the work conducted by Olekseyuk \textit{et.al}, whose experimental value is 1.69 eV but overestimates the theoretical value $\sim$ 0.9 eV.\cite{Olekseyuk2005}

\par The DOS refers to the number of electronic states attainable per unit energy range for electrons in a material.\cite{Kukreti2021} The calculated DOS for the pristine structure of Cu$_2$XSiS$_4$ (X = Ge, Sn, Pb) are shown in Figure \ref{Fig.3}, revealing the contribution of atoms in the valence and conduction band. The Fermi energy has been set at 0 eV for all the plots, and the DOS amplitude within the range of 0 to 60 States/eV was taken. For X = Ge, the contribution near the Fermi region in the conduction band is mainly comprised of the Ge atom, followed by a slightly lesser contribution from the S atom, with both increasing as the energy increases. A negligible contribution of Si and Cu atoms can also be observed in the region of 2 to 3 eV. Conversely, in the valence band, the Cu and S atoms contribute equally near the Fermi region, with the Cu atom contribution increasing in the deep level region of the valence band. A minor contribution of Ge and Si atoms can also be observed in the region of -2 to -3 eV. Similarly, in the valence band of the X = Sn compound, the region near the Fermi level is composed of an equal contribution of Cu and S atoms. However, the Cu atom contribution in the deep level region rises prominently, with an amplitude exceeding 45 States/eV. In the conduction band, the Sn atom contributes the most near the Fermi region, followed by S atoms, and a negligible contribution of the Si atom. The Sn atom contribution decreases as the energy increases. Furthermore, for X = Pb, the contributions of each atom in the conduction band are similar to those of the X = Sn compound. The valence band also follows the same trend as the X = Sn compound, as mentioned earlier, with Cu and S atoms mostly contributing in the region near the Fermi level and in the deep level region, the Cu atom contribution increases significantly while the S atom contribution decreases.

\subsection{Optical properties} The optical characteristics of materials, governed directly by their ground-state electronic structure, determine how they interact with light and other types of electromagnetic radiation.\cite{Persson2015} It plays a critical role in assessing a material's applicability in optoelectronic and photovoltaic devices. The analysis of optical properties is typically carried out using the complex dielectric function, absorption coefficient, and refractive index. The optical characteristics in this study were calculated up to 8 eV photon energies to reveal the response of Cu$_2$XSiS$_4$ (X = Ge, Sn, and Pb) to solar and high-energy wave radiations. The observed optical constants have been plotted along the \textit{xx}-, \textit{yy}-, and \textit{zz}-axes, and the corresponding optical responses along the \textit{xx}- and \textit{yy}-axes are identical and overlap due to the isotropy of planar symmetry.\cite{Chen2019} However, due to anisotropic symmetry, the response along the \textit{zz}-axis differed.\cite{Brik2014}

\par
\begin{figure*}[hbt!]
	\centering
	\includegraphics[height=15cm]{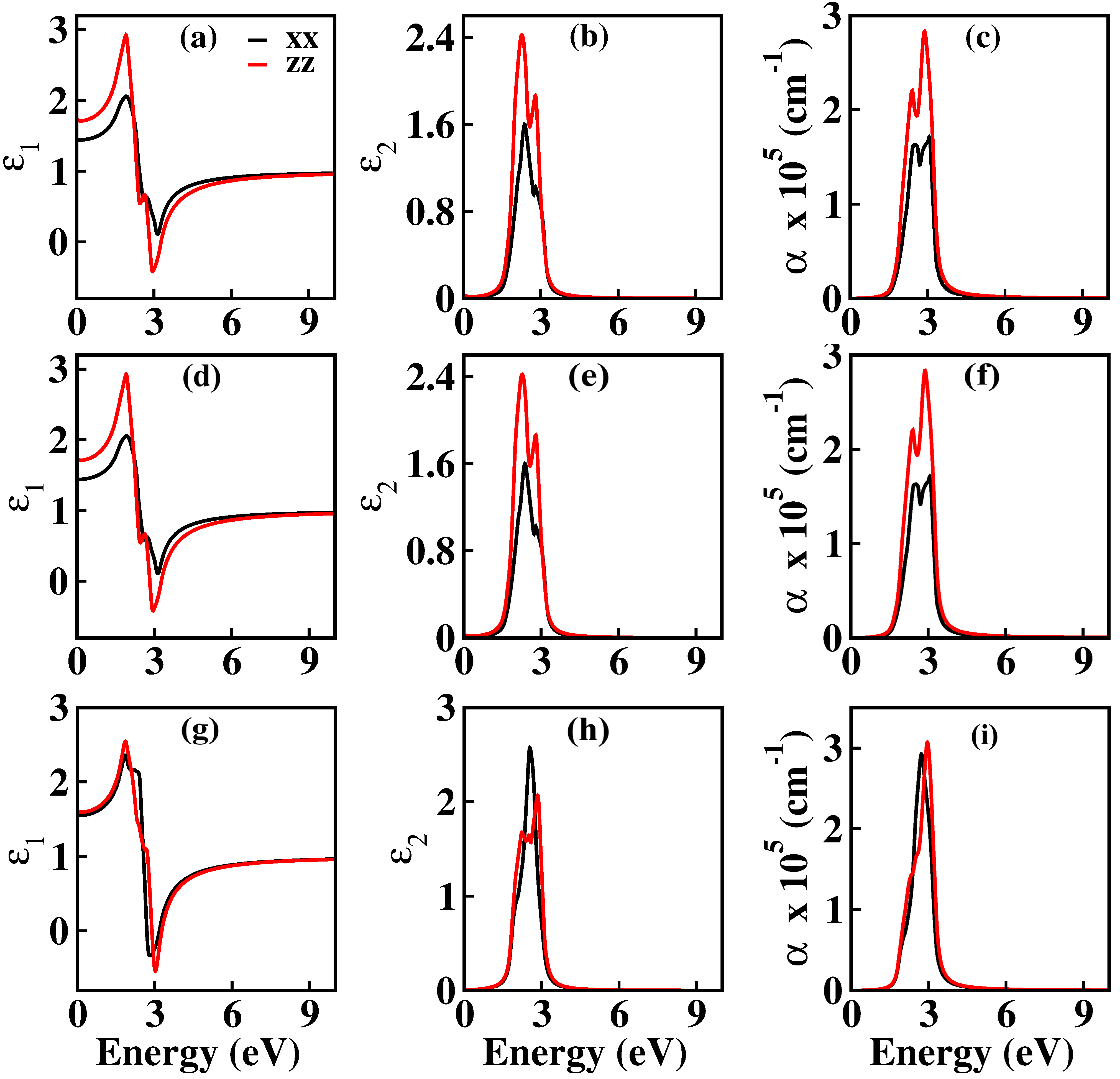}
	\caption{Calculated optical properties ($\epsilon_1$, $\epsilon_2$, and $\alpha$) of the studied compounds. Figure (a-c) for X = Ge, Figure (d-f) for X = Sn and Figure (g-i) for X = Pb}
	\label{Fig.4 Optical}
\end{figure*}

 The dielectric function describes how a material responds to variations in charge distribution, particularly under the influence of an external electric field. When photons interact with the electric field, electrons transition from occupied energy states to unoccupied energy states.\cite{SalmanKhan2024} Such a response of an electron can be well studied in terms of its complex dielectric functions given by \cite{Srivastava2020}
\begin{equation}
	\epsilon = \epsilon_1(\omega) + i\epsilon_2(\omega)
	\label{Eq 2}
\end{equation}
where $\epsilon_1$ and $\epsilon_2$ are the real and imaginary parts of the dielectric function, respectively. From the knowledge of the real and imaginary parts of the dielectric tensor, it is possible to predict other important optical properties such as the absorption coefficient and refractive indices.\cite{Ghosh2024} To understand how a material absorbs light, it is necessary to know about the imaginary part of the dielectric function. The absorption of a material is significant when the absorptive component of the electronic dielectric function $\epsilon_2$ has a large value. The imaginary part of the dielectric function is obtained from the linear response in the long wavelength limit, taking the form \cite{Ziane2021}
\begin{equation}
		\epsilon_2(\omega) = \frac{e_2h}{\pi m^2 \omega^2}\sum_{c,v} \int_{BZ}|M_{ev}(k)|^2\delta[\omega_{cv}-\omega]d^3k
		\label{Eq 3}
\end{equation}
The real part can be obtained from the imaginary part by using the Kramers-Kronig transformation, which is given by \cite{Camps2012}
\begin{equation}
	\epsilon_1 = 1 + \frac{2}{\pi}p\int_{0}^{\infty}\frac{\omega\epsilon_2(\omega)}{\omega'^2-\omega^2}d\omega'
	\label{Eq 4}
\end{equation}
The dielectric functions of the real and imaginary parts, as well as the absorption coefficient, are shown in Figure \ref{Fig.4 Optical}. In the real part of the dielectric constant, the X = Ge compound exhibits the maximum value along the \textit{zz}-axis at 2.91 arbitrary units (a.u) at 1.99 eV. Moreover, both X = Sn and Pb compounds also show peaks along the \textit{zz}-axis as well at 1.99 a.u and 1.86 a.u in the energy of 2.91 eV and 2.59 eV, respectively. The imaginary part of the dielectric function for X = Ge exhibit a maximum value along the \textit{zz}-axis at 2.40 a.u with energy 2.41 eV. Similarly, for X = Sn, the peak is observed at 2.35 a.u with energy 2.40 eV. Meanwhile, the X = Pb compound shows maximum value along the \textit{xx}-axis contrary to all other results, at 2.62 a.u with the energy 2.55 eV. The calculated static dielectric constant for the real part is shown in Table \ref{Table 2} along the \textit{xx} and \textit{zz}-axes. 

\begin{table}[hbt!]
	\small
	\caption{Calculated static real $\epsilon_1$(0) part of the dielectric constant and static refractive indices ($n$) along the \textit{xx}- and \textit{zz}-axis for Cu$_2$XSiS$_4$ (X = Ge, Sn, Pb).}
	\label{Table 2}
	\begin{tabular*}{0.48\textwidth}{@{\extracolsep{\fill}}l|ll}
		\hline
		X & \ \ \  Static $\epsilon_1$(0)&  \ \ \ Static $n$ \\
		\hline
        & \ \ \ \ \ \textit{xx}, \textit{zz} & \ \ \ \ \ \textit{xx}, \textit{zz} \\
        \hline
        Ge&\ \ \ 1.73, 1.43& \ \ \ 1.19, 1.31 \\
        Sn&\ \ \ 1.60, 1.60& \ \ \ 1.25, 1.27 \\
        Pb&\ \ \ 1.59, 1.55& \ \ \ 1.24, 1.26 \\
		\hline
	\end{tabular*}
\end{table}

\par To validate the findings of the dielectric function, we have examined the absorption coefficient, which conveys information regarding the attenuation of light intensity per unit distance within the medium. The last row of Figure \ref{Fig.4 Optical} represents the absorption coefficients for Cu$_2$XSiS$_4$ (X= Ge, Sn, Pb) in the energy range of 0 to 10 eV along the \textit{xx} and \textit{zz}-axis. It is clear from the observed plot that all the compounds' absorption lies within the visible-UV region. We can see that the highest peak occur at 2.83 x 10$^5$ cm$^{-1}$, 2.85 x 10$^5$ cm$^{-1}$, and 3.05 x 10$^5$ cm$^{-1}$ for X = Ge, Sn and Pb respectively. Therefore, we can say that the absorption peak shifted towards a higher value when the ionic radii of the substituent divalent cation site increase down the group. A steep onset of absorption can be seen for each compound, which suggests efficient light absorption. The onset starts at 1.49 eV, 1.45 eV, and 1.54 eV for X = Ge, Sn, and Pb, respectively.  
\par
\begin{figure*}[hbt!]
	\centering
	\includegraphics[height=6cm]{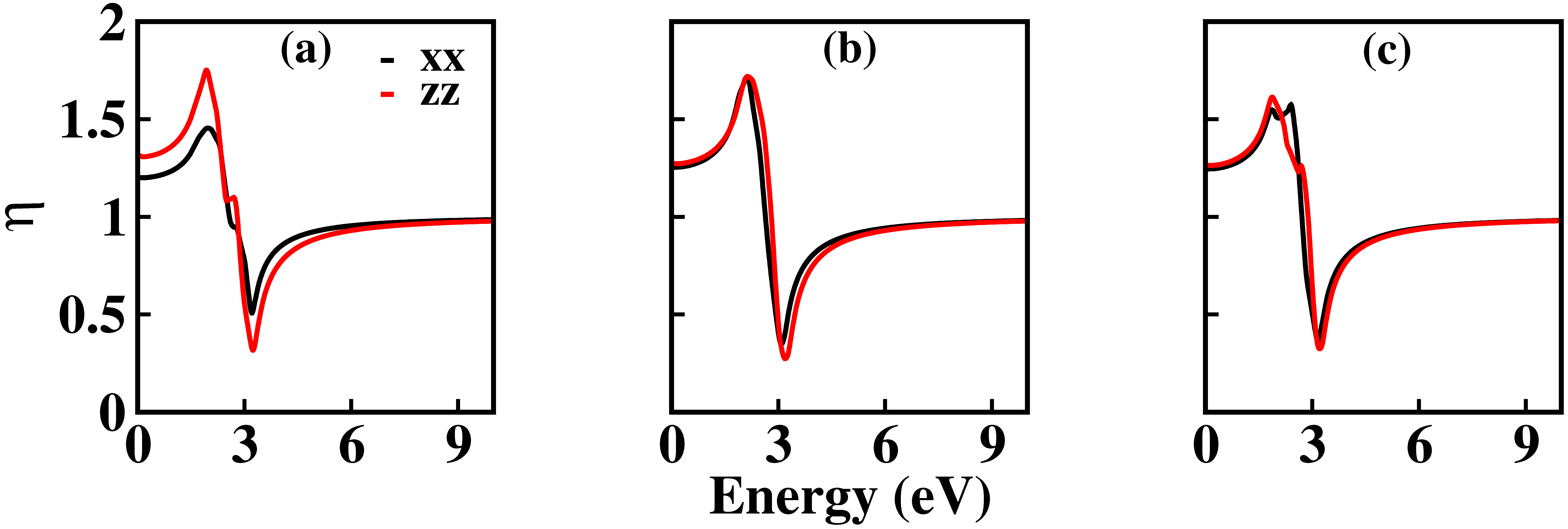}
	\caption{Refractive indices of Cu$_2$XSiS$_4$. (a) for X=Ge, (b) for X=Sn, (c) for X=Pb}
	\label{Fig.5 Optical}
\end{figure*}

 The refractive index spectra, a crucial property which describes the propagation of light through a material, are shown in Figure \ref{Fig.5 Optical}. The calculated values of the refractive index are found to be 1.44 a.u and 1.74 a.u along the \textit{xx} and \textit{zz}-axis, respectively, for X = Ge at energy 2.04 eV and 1.99 eV. For X = Sn, the peaks along the \textit{xx} and {zz}-axis are the same with a value of 1.70 a.u at the same energy of 2.17 eV. Likewise, the absorption value for X = Pb along the \textit{xx} and \textit{zz}-axis values are nearly identical at around $\sim$ 1.60 a.u, but within different energies of 2.43 eV and 1.94 eV, respectively. The static reflective indices for all the compounds are shown in Table \ref{Table 2}. 

\subsection{Mechanical properties} To determine the compounds' stability, we determined the elastic constants as it is crucial for understanding their physical properties. Herein, we calculated the elastic constants using the finite-strain theory.\cite{Kiely2021} The mechanical stability criteria, also known as Born criteria for the trigonal system, are as follows:\cite{Hossain2019}
\begin{equation}
	\begin{split}
	C_{11} - C_{12}>0 \\
	(C_{11}+C_{12})C_{33}-2C^2_{14}>0 \\
	(C_{11}-C_12)C_{44}-2C^2_{14}>0 \\
	\end{split}
	\label{Eq 5}
\end{equation}

\begin{table*}[hbt!]
\small
	\caption{Calculated elastic constants (in GPa), Cauchy Pressure (in GPa), and Kleimann coefficient (unitless)}
	\label{Table 3}\renewcommand{\arraystretch}{1.25}
	\begin{tabular*}{\textwidth}{@{\extracolsep{\fill}}llllllllllll}
		\hline
		X & $C_{11}$ & $C_{12}$ & $C_{13}$ & $C_{22}$& $C_{33}$ & $C_{44}$ & $C_{55}$ & $C_{66}$ & $C_{12}$ - $C_{44}$ & $\zeta$ & \\
		\hline		Ge&72.20&33.86&41.63&71.70&84.38&22.84&22.33&18.05&11.02&0.59& \\
Sn&90.02&35.72&40.21&90.35&96.55&22.33&21.72&26.15&13.32&0.53& \\
Pb&82.97&31.93&38.85&76.59&91.31&19.44&22.76&21.63&12.49&0.52& \\
\hline
	\end{tabular*}
\end{table*}

\begin{table*}
\small
	\caption{Calculated values of elastic moduli $-$ Bulk modulus(B), Young's modulus (E), Shear modulus (G) all in GPa units, Pugh's ratio (k), and Poisson's ratio (\textit{v})(unitless).}
	\label{Table 4}

\begin{tabular*}{\textwidth}{@{\extracolsep{\fill}}llllll}

		\hline
		X & B & E & G & \textit{k} & \textit{v} \\
		\hline
		Ge & 50.53 & 52.44 & 19.76 & 2.55 & 0.32  \\
		Sn & 50.74 & 58.35 & 22.30 & 2.27 & 0.30  \\
	    Pb & 56.28 & 64.58 & 24.67 & 2.28 & 0.30  \\
		\hline
	\end{tabular*}
\end{table*}
All the elastic constants (see Table \ref{Table 3}) for our examined materials meet the criteria listed above. We found that the constants $C_{11}$, $C_{22}$, and $C_{33}$ are significantly higher than $C_{44}$, $C_{55}$, and $C_{66}$. Thus, the compound under investigation appears to be more resistant to axial compression than shear deformation, as evidenced by the fact that the bulk modulus (k) is greater than the shear modulus.\cite{DeJong2015} Larger constant values, such as $C_{11}$, $C_{22}$, and $C_{33}$, indicate that the compound exhibits extremely anisotropic single-crystal elasticity.\cite{Mohapatra2008} To investigate internal deformation stability, we derived the Kleinman coefficient using the following equation.\cite{Celestine2024}
\begin{equation}
	\zeta = \frac{C_{11}+8C_{12}}{7C_11+2C_{12}}
    \label{Eq.6}
\end{equation}
where 0 $<$ $\zeta$ $>$1 . This coefficient ranges from 0 to 1, which elucidates whether the major contribution is towards bond stretching or bond bending, depending on the value of $\zeta$ in each compound.\cite{Naher2021} From Table \ref{Table 3}, we can see that for each compound, the major contribution can be towards both bending and stretching. 

\par The mechanical stability and elastic behaviour were evaluated through the calculation of their macroscopic elastic moduli, including bulk modulus (B), Young's modulus (E), shear modulus (G), Pugh's ratio (\textit{k}), and Poisson's ratio (\textit{v}) (see Table \ref{Table 4}). These parameters, derived from the elastic stiffness constants, offer critical insight into the structural integrity and mechanical reliability of the materials.\cite{Boucetta2025} The bulk modulus values, ranging from 50.53 to 56.28 GPa, indicate that all compositions possess a moderate degree of resistance to uniform volumetric compression, with the Pb-containing structure demonstrating the highest incompressibility. Young's modulus, which reflects the stiffness of a material under uniaxial stress, increases progressively from 52.44 GPa (Ge) to 64.58 GPa(Pb), implying that Pb substitution marginally enhances the rigidity of the lattice. Conversely, the shear modulus values are relatively low across all compositions, suggesting that these materials are intrinsically soft and prone to shear deformation under applied stress.\cite{Kahlaoui2020}

\par To further assess ductility, Pugh's ratio was calculated and found to be significantly greater than the critical threshold of 1.75 for all three materials: approximately 2.55 for Ge, 2.27 for Sn, and 2.28 for Pb. These values indicate a strong tendency towards ductile behaviour, which is advantageous for practical applications where resistance to fracture and mechanical failure is required.\cite{Zhao2017} Additionally, the calculated Poisson's ratio, ranging from 0.30 to 0.32, fall well within the accepted bounds (-1.0$<$\textit{v}$<$0.5)\cite{Chi2024} for mechanically stable crystalline solids, and is indicative of a balanced response between longitudinal and transverse strain under stress.\cite{Greaves2001} Collectively, these results not only affirm the mechanical stability of compounds but also highlight their potential suitability for use in optoelectronic or thin-film device applications where a combination of flexibility, mechanical resilience, and ductility is desirable.  

\begin{figure*}
    \centering
    \includegraphics[height=6.5cm]{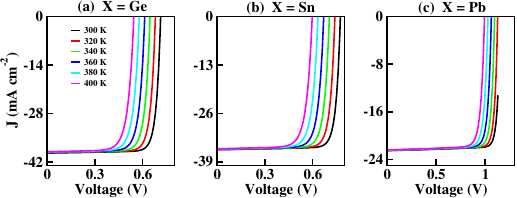}
    \caption{J-V characteristic curve of a solar cell with varying temperature (a) X = Ge (b) X = Sn (c) X = Pb}
    \label{Fig.6}
\end{figure*}

\begin{figure*}
    \centering
    \includegraphics[height=10cm]{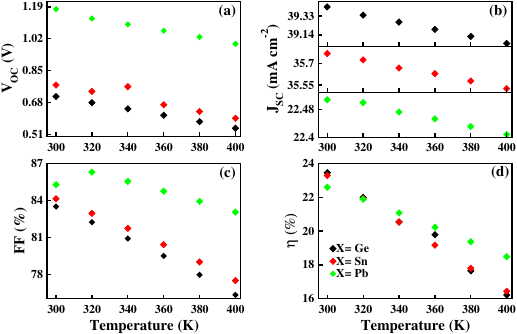}
    \caption{Relationship between temperature and solar cell parameters}
    \label{Fig.7}
\end{figure*}

\section{Solar--cell Simulation} 
\label{scaps}
\subsection{Device architecture and material parameters} Solar Cell Capacitance Simulator-One Dimension (SCAPS-1D) is an advanced mathematical and physical model. It is an excellent tool for simulating the performance of thin-film solar cells, developed by the University of Gent, Belgium.\cite{Burgelman2000, Baro2023} The principle of this software is based on theoretical and computational resolution of Poisson's equation and electron and hole continuity equations. Poisson's equation relating to the space charge density and the electric field generated by the p-n junction can be expressed as\cite{Medina2023}
\begin{equation}
\begin{split}
    \frac{\partial^{2}\psi}{\partial{x^2}}+\frac{q}{\epsilon_r\epsilon_0}[p(x)-n(x)+N_D^+(x) \\
    -N_A^-(x)+p_t(x)-n_t(x)] = 0
\end{split}
\label{Eq.7}
\end{equation}
where $\psi$ is the electrostatic potential; q is the electron charge; N$_D^+$ and N$_A^-$ are the ionized donor and acceptor densities, respectively; p and n are the electron and hole densities, respectively; $\epsilon_r$ and $\epsilon_0$ are respectively the vacuum and relative permittivity; p$_t$ represent trapped holes; n$_t$ represents trapped electrons; and x denotes the position of electrons in the x-coordinate.
\par The steady-state continuity equation for electrons and holesis\cite{HOUIMI2021}
\begin{equation}
    \frac{\partial J_n}{\partial x}+G-R=0
    \label{Eq 8}
\end{equation}
\begin{equation}
    \frac{\partial J_p}{\partial x}+G-R=0
    \label{Eq 9}
\end{equation}
Here, J$_n$ and J$_p$ represent electron and hole current densities, respectively; G represents the carrier generation rate; and R is the net recombination from direct and indirect recombination.

\begin{figure*}[ht]
    \centering
    \includegraphics[height=6.6cm]{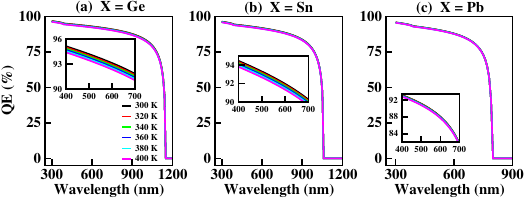}
    \caption{Relationship between temperature and solar cell quantum efficiency}
    \label{Fig.8}
\end{figure*}

\par The presence of an electric current or a carrier concentration gradient causes the movement of electrons or holes, generating a current in a semiconductor material. The electron and hole current densities are\cite{KANNAN2023}
\begin{equation}
    J_n=qn\mu_nE+qD_n \frac{\partial n}{\partial x}
    \label{Eq 10}
\end{equation}
\begin{equation}
    J_p=qp\mu_nE-qD_p \frac{\partial p}{\partial x}
    \label{Eq 11}
\end{equation}
where $\mu_n$ and $\mu_p$ are electron and hole mobilities, D$_n$ and D$_p$ are electron and hole diffusion coefficients, respectively, and E is the electric field.
\par The architecture of Charge-Transport material-based n-i-p planar solar cell structure is shown in Figure \ref{Fig.1 Structures}. The layers composing our heterojunction solar cell are as follows: an n-TiO$_2$ thin film for the electron transport layer, quaternary chalcogenides Cu$_2X$SiS$_4$ (X = Ge, Sn, Pb) as an active absorbing layer, and p-CuO thin-film as the hole transport layer. The transparent conductive oxide (FTO) serves as the front contact, and gold (Au) acts as the back contact. The output of this computation includes quantum efficiency (QE), current-voltage (I-V) characteristics such as open-circuit voltage (V$_{oc}$), short-circuit current density (J$_{sc}$), fill factor (FF), and efficiency ($\eta$). All the necessary material parameters are calculated using DFT, and some are borrowed from the literature. These input parameters are listed in Table (\ref{Table 5} and \ref{Table6}).  

\subsection{Effect of operating temperature} 
In general, the conventional operating temperature is typically taken as 300K. However, given that sunlight conditions vary throughout the day and across seasons, it is essential to analyse the performance of the solar cell device under different temperature conditions. This allows for more accurate modelling and ensures reliable operation of the device in real-world environments. To analyse the effect of temperature on the proposed device, the operating temperature was varied from 300 to 400 K. All the other input parameters were kept constant.
\par The J-V characteristic curve is shown in Figure.\ref{Fig.6} and their corresponding solar cell parameters for each examined compound are shown in Figure.\ref{Fig.7}. It can be observed from the J-V curve that the best behaviour is found at 300K. V$_{oc}$ versus operating temperature is shown in Figure.\ref{Fig.7}(a). Generally, solar cells are negatively influenced by temperature, and their PCE decreases as temperature increases. X = Pb compounds exhibit the highest V$_{oc}$, followed by Sn, and finally Ge-based compounds, where all three compounds decrease as the temperature increases. For X = Ge compound, V$_{oc}$ has a small decrease of 0.03V when the temperature increases, going from 0.71V at 300K to 0.54V at 400K. Similarly, for X = Sn, the V$_{oc}$ decreases from 0.77V to 0.59V at 300K to 400K, respectively, with a decreasing rate of 0.03V. The X = Pb compound starts at 1.17 V at 300K, then decreases at the same rate and stops at 0.99 V at 400 K. This behaviour can be understood by the following equation\cite{ADEWOYIN2019}
\begin{equation}
    {V_{oc}}=\frac{kT}{q} ln (\frac{J_{sc}}{J_0}+1)
    \label{eq 12}
\end{equation}
where (kT/q) is a thermal voltage, J$_{sc}$ is short-circuit current density, and J$_0$ is the saturation current density which can be expressed as \cite{MAHARANA2022}
\begin{equation}
    J_0 = \frac{qD_pp_{n0}}{L_p}+\frac{qD_nn_{p0}}{L_n}
    \label{eq 13}
\end{equation}
where L$_p$ and L$_n$ are diffusion lengths of holes and electrons, respectively, D$_p$ and D$_n$ are the diffusion coefficients for holes and electrons, respectively, p$_{n0}$ is the equilibrium hole density in region n, and n$_{p0}$ is the equilibrium density of electrons in the p region. A temperature rise directly causes an exponential growth in reverse saturation current, which is a primary factor in reducing solar cell parameters.
\par Figure.\ref{Fig.7}(b) shows that the short-circuit current also decreases with an increase in temperature. For X = Ge, the J$_{sc}$ reaches up to 39.42V at 300K, which gradually decreases at a rate of 0.03V to 39.05V at 400K. The X = Sn compound decreases at a slightly slower rate of 0.05V, starting from 35.77V to 35.52V at 300K and 400K, respectively. For X = Pb, the J$_{sc}$ change at a rate of 0.02V from 22.50V at 300K to 22.40V at 400K. From Eq.\ref{eq 12}, it can be seen that the V$_{oc}$ is logarithmically proportional to the ratio of J$_{sc}$ to J$_0$. Thus, the observed lower J$_{sc}$ with higher temperature explains the same trend observed in the V$_{oc}$ plot. 
\par In Figure.\ref{Fig.7}(c), the fill factor (FF) for each studied compound exceeds 80\% at room temperature, which is ideal for a solar cell as it indicates a more efficient cell with less internal resistance. The FF behaviour can be understood by the following equation\cite{BENZETTA2021}
\begin{equation}
    FF=\frac{P_m}{J_{sc}V_{oc}}
    \label{eq 14}
\end{equation}
where P$_m$ is the maximum possible power output. The FF for X = Ge compounds reach up to 83.50\% at 300K, then decrease to 76.31\% at 400K. Similarly, the FF of X = Sn compound decreases from 84.13\% at 300K to 77.48\% at 400 K. The X = Pb compound exhibits the highest FF among the three studied compounds. At room temperature, the FF is 85.28\%, then slightly increases when the temperature rises to 320K at a value of 86.30\%, and gradually decreases when the operating temperature is further increased.

\begin{table*}[ht]
\small
	\caption{Parameters for the different layers of the proposed solar cells with Cu$_2$XSiS$_4$(X=Ge, Sn, Pb) as absorber layer}
	\label{Table 5}
\begin{tabular*}{\textwidth}{@{\extracolsep{\fill}}|l|l|lll|l|}
		\hline
Parameters & CuO\cite{Lofty2025} & X = Ge & X = Sn & X = Pb & TiO$_2$\cite{Lofty2025} \\
		\hline
Thickness (nm)& 50 & 1500 & 1500 & 1500 & 50  \\
Dielectric Permittivity& 18.10 & 9.0 & 9.0 & 9.0 & 10  \\
Bandgap (eV)& 1.2 & 1.08 & 1.17 & 1.56 & 3.2 \\
Electron affinity (eV) & 4.07 & 4.1 & 4.1 & 4.1 & 4.2 \\
Effective density of states of VBM (cm$^{-3}$) & 5.5 $\times$ 10$^{20}$ & 7.31 $\times$ 10$^{19}$ & 1.23 $\times$ 10$^{20}$ & 6.36 x 10$^{19}$& 6 $\times$ 10$^{17}$\\
Effective density of states of CBM (cm$^{-3}$) & 2.2 $\times$ 10$^{19}$ & 8.61 $\times$ 10$^{18}$ & 2.97 $\times$ 10$^{19}$& 2.73 $\times$ 10$^{19}$& 2 $\times$ 10$^{17}$ \\
Acceptor concentration (cm$^{-3}$) & 1.0 $\times$ 10$^{18}$ & 1.0 $\times$ 10$^{19}$ & 1.0 $\times$ 10$^{19}$  & 1.0 $\times$ 10$^{19}$  & 0 \\
Donor concentration (cm$^{-3}$) & 0 & 0 & 0 & 0 & 1.0 $\times$ 10$^{17}$ \\
Mobility of hole (cm$^{2}$/V.s) & 20 & 35.89 & 15.70 & 16.59 & 250 \\
Mobility of electron (cm$^{2}$/V.s) & 200 & 8.62 & 6.08 & 9.45 & 100 \\
Electron thermal velocity (cms$^{-1}$) & 10$^7$ & 1.6 $\times$ 10$^5$ & 1.10 x 10$^5$&1.13 $\times$ 10$^5$ & 10$^7$ \\
Hole thermal velocity (cms$^{-1}$) & 4.6 $\times$ 10$^6$ & 8.17 $\times$ 10$^4$ & 6.87 x 10$^4$ & 8.56 $\times$ 10$^4$ & 10$^7$  \\
Defect density N$_t$ (cm$^{-3}$) & 1$\times$10$^{15}$ & 1 $\times$ 10$^{15}$ & 1 $\times$ 10$^{15}$ & 1 $\times$ 10$^{15}$ &1 $\times$ 10$^{15}$  \\
		\hline
	\end{tabular*}
\end{table*}

\begin{table*}[ht]
\small
	\caption{The front and back contact parameters are taken from the previous work\cite{Mahmoud, Atowar2021, Lofty2025}}
	\label{Table6}
\begin{tabular*}{\textwidth}{@{\extracolsep{\fill}}|l|l|l|}
		\hline
		Contacts &
Front metal contact (FTO) &
Back metal contact (Au) \\
			\hline
Metal work function (eV) & 4.07 & 4.98\\
Surface recombination velocity of electrons (cm/s)&
1.000 $\times$ 10$^7$ &
1.000 $\times$ 10$^7$\\
Surface recombination velocity of hole (cm/s)&1.000 $\times$ 10$^7$&
1.000 $\times$ 10$^7$\\
\hline
	\end{tabular*}
\end{table*}

\par It can be observed from Figure \ref{Fig.7}(d) that the efficiency for X = Pb decays at a slower rate as compared to X = Ge and Sn compounds when the temperature rises. The PCE($\eta$) of a solar cell can be calculated using\cite{MOUSTAFA2022}
\begin{equation}
    \eta=\frac{J_{sc}V_{oc}FF}{P_{in}}
    \label{eq 15}
\end{equation}
where P$_{in}$ is the total power from sunlight incident on the cell. 
\par The X = Ge exhibits the highest efficiency of 23.46\% at 300K. However, it decays faster compared to the other studied compounds, with 16.22\% at 400K. The efficiency of the X = Sn compound behaves similarly, with an efficiency of 23.29\% at 300K. As the temperature rises, the efficiency decreases to 16.44\% at 400K. The X = Pb efficiency is 22.60\%, which is the lowest among the three at 300K, but at 400 K, an efficiency of 18.49\% is observed, which is the highest. This implies that the X = Pb compound can withstand temperature variation the best than the other studied compounds. Therefore, we can say that X = Ge compound is ideal for standard conditions, and the X = Pb compound shows potential for long-term stability under real-world conditions. The quantum efficiency (QE) of a solar cell demonstrates the device's capacity to convert light into electricity. The relationship between operating temperature and solar cell quantum efficiency is graphically presented in Figure.\ref{Fig.8}. The X = Ge and Sn compounds exhibit similar QE behaviour, as shown in Figure.\ref{Fig.8}(a and b), with a remarkably high QE of above 90\% from 300 to 800 nm, indicating that they can absorb light radiation in both the visible and infrared regions. A gradual decrease can be seen at first, a sharp decline at $\sim$1100 nm, and then finally reaches zero beyond. Meanwhile, the X = Ge compound has QE of over 90\% from 300-600 nm, with a sharp decline at $\sim$800 nm and no measurable response at longer wavelengths as shown in Figure.\ref{Fig.8}(c). A moderate decrease can be observed when the operating temperature increases for each compound. The X = Ge and Sn compounds demonstrate a comparable response to the temperature variation, while the X = Pb compound has a minimal change across the tested temperature range.

\section{Conclusion} In this work, we thoroughly investigated the structure, electronics, optical, and mechanical properties of Cu$_2$XSiS$_4$ (X = Ge, Sn, and Pb) quaternary chalcogenide materials based on DFT calculations. The stability of these compounds was validated using formation energy, Born stability criteria, and AIMD simulation. We also observed that all the compounds exhibit an indirect band gap, with a clear trend of increasing band gap down the group. As the imaginary part of the dielectric constants aligns well with the absorption coefficients, all the $\alpha$-peaks fall under the vis-UV region for all three compounds, indicating their suitability in photovoltaic applications. Additionally, using SCAPS-1D software, we have modeled and assessed a FTO/TiO$_2$/Cu$_2$XSiS$_4$/CuO/Au heterojunction solar cell. This study also investigated the effect of operating temperature on the performance of the absorber layer. 
By optimizing the parameters, we have found that the V$_{oc}$ increases when moving from Ge to Sn to Pb. Similarly, the FF also increase (Ge $\rightarrow$ Pb) in the group, while the J$_{sc}$ follows the opposite trend. The results indicate that the Cu$_2$PbSiS$_4$ compound can best withstand temperature variation. However, Cu$_2$PbSiS$_4$ exhibits the highest PCE among the three compounds at room temperature. The recorded room temperature PCEs are 23.46\%, 23.29\% and 22.60\% for Cu$_2$GeSiS$_4$, Cu$_2$SnSiS$_4$, and Cu$_2$PbSiS$_4$, respectively.   We have calculated the values of all the parameters and reported them for the first time. Therefore, all the calculated results are novel and promising for further experimental validation and application in PV solar cells.

\section{Acknowledgments}
DPR acknowledges Anusandhan National Research Foundation (ANRF), a statutory body of the Department of Science \& Technology (DST), Government of India, for the project Sanction Order No.: CRG/2023/000310, dated: 10 October 2024.\\  DPR and KCB also gratefully acknowledge the computational resources provided by the \underline{Paramrudra High-Performance Computing (HPC)} facility at the \underline{Inter-University Accelerator Centre (IUAC)}, New Delhi.

\section*{Author contributions}
\textbf{H. Laltlanmawii:} Formal analysis, Visualisation, Validation, Literature review, Performed calculation, Writing-original draft, writing-review \& editing.\\
\textbf{L. Celestine:} Formal analysis, Visualisation, Validation, Literature review, Performed Calculation, writing-review \& editing.\\
\textbf{R. Zosiamliana:} Formal analysis, Visualisation, Validation, writing-review \& editing. \\
\textbf{B. Chettri:} Formal analysis, Visualisation, Validation, SCAPS-1D calculation, writing-review \& editing. \\
\textbf{S. Bhattarai:} Formal analysis, Visualisation, Validation,  SCAPS-1D calculation, writing-review \& editing. \\
\textbf{K. C. Bhamu:} Formal analysis, Visualisation, Validation, writing-review \& editing. \\
\textbf{D. P. Rai:} Project management, Supervision, Resources, software, Formal analysis, Visualisation, Validation, SCAPS-1D calculation, writing-review \& editing. 

\section*{Conflicts of interest}
There are no conflicts to declare.

\section*{Data availability}
The data that support the findings of this study are available from the corresponding author upon reasonable request.

\nocite{*}
\bibliography{Hani}

\end{document}